\documentclass[10pt,a4paper,twoside]{article}
\usepackage{epsfig}
\usepackage{baltlat5}
\usepackage{wrapfig}
\begin{document}
\ \
\vspace{0.5mm}

\setcounter{page}{1}
\vspace{5mm}

\titlehead{Baltic Astronomy, vol.\ts 14, XXX--XXX, 2005.}

\titleb{NLTE ANALYSES OF SDB STARS: PROGRESS AND PROSPECTS}


\begin{authorl}
\authorb{N.~Przybilla}{1}
\authorb{M.~F.~Nieva}{1,2} and
\authorb{H.~Edelmann}{1}
\end{authorl}

\begin{addressl}
\addressb{1}{Dr.\,Remeis-Sternwarte Bamberg, Sternwartstr.\,7, D-96049 Bamberg, Germany}

\addressb{2}{Observat\'orio Nacional, R. Gal. Jos\'e Cristino 77, 20921-400,
S\~ao Crist\'ov\~ao, Rio de Janeiro, RJ, Brasil}

\end{addressl}

%

\submitb{Received 2005 July 31}

\begin{abstract}
We report on preliminary results of a hybrid non-LTE analysis of
high-resolution, high-S/N spectra of the helium-rich subdwarf B star 
Feige\,49 and the helium-poor sdB HD\,205805. Non-LTE effects are found to have 
a notable impact on the stellar parameter and abundance determination. In particular
the He\,{\sc i} lines show significant deviations from detailed balance, with the
computed equivalent widths strengthened by up to $\sim$35\%. Non-LTE abundance 
corrections for the metals (C, N, O, Mg, S) are of the order 
$\sim$0.05\,--\,0.25\,dex on the mean, while corrections of 
up to $\sim$0.7\,dex are derived for individual transitions. The non-LTE
approach reduces systematic trends and the statistical uncertainties in the
abundance determination. Consequently, non-LTE analyses of a larger sample of objects have
the potential to put much tighter constraints on the formation history of the different 
sdB populations than currently discussed.
\end{abstract}

\vskip1mm


\resthead{Non-LTE analyses of sdB-stars: progress and prospects}{Przybilla, Nieva \& Edelmann}


\sectionb{1}{INTRODUCTION}

Astrophysics faces the general difficulty that almost all 
relevant information cannot be directly inferred from observation. 
One has to rely on the interpretation of the radiation from a light-emitting 
plasma and its interaction with matter. Accurate physical modelling is crucial, 
with systematic uncertainties often dominating the error budget at present. 
In the case of subdwarf B (sdB) stars quantitative spectroscopy is well established, 
relying on grids of either line-blanketed LTE or metal-free non-LTE model
atmospheres (e.g. Heber \& Edelmann~2004, and references therein). Here, we 
investigate what improvements on sdB analyses can be expected from a hybrid 
non-LTE method, which avoids the weaknesses of both traditional approaches by
combining metal line-blanketed model atmospheres with non-LTE line formation. 
The impact on the stellar parameter determination is of special interest, as
these provide the basis for all further interpretation. Then, in the second part, 
non-LTE metal abundances are determined from observations in the visual spectral 
range for the first time. 
More accurate stellar parameter and abundance determinations can be expected
to shed further light on the formation mechanisms of sdB stars that have been debated over
the last decades.

\sectionb{2}{MODEL CALCULATIONS \& OBSERVATIONAL DATA}

The model calculations are carried out in a hybrid approach, thus solving
the (so-called) restricted non-LTE problem. Hydrostatic,
plane-parallel and line-blanketed -- via an opacity sampling (OS) technique -- LTE
model atmospheres are computed with the {\sc Atlas12} code
(Kurucz~1996), which is in particular suited for the analysis of
chemically peculiar stars. Then, non-LTE line formation is performed on the
resulting model stratifications. The coupled radiative transfer and statistical 
equilibrium equations are solved and spectrum synthesis with refined line-broadening 
theories is performed using {\sc Detail} and {\sc Surface} (Giddings~1981;
Butler \& Giddings~1985). Both codes have undergone major revisions
and improvements over the past few years. State-of-the-art non-LTE model atoms
(see Table~1) are utilised for the stellar parameter and abundance determination.

The model atoms are largely based on data from quantum-mechanical {\it ab-initio} 
computations using the $R$-matrix method in the close-coupling approximation, 
which typically have uncertainties of the order $\sim$10--20\%. Data sources 
comprise the astrophysically motivated Opacity and IRON Project, but also the 
vast physics literature, see the original publications for details. 
This allows for a realistic treatment not only of the (non-local) radiative 
processes, which drive the departures from detailed balance. Also the 
thermalising collisions are accurately represented for the relevant transitions, 
in contrast to the usual approach of applying simple approximation formulae, 
which can be in error by orders of magnitude. Finally, state-of-the-art
line-broadening theories are accounted for, like the data of Stehl\'e \&
Hutcheon~(1999) for the Stark broadening of the hydrogen lines.

Spectra of HD\,205805 and Feige\,49 were taken with {\sc Feros} (Kaufer et
al.~1999) on the ESO\,1.52m and 2.2m telescopes, respectively. The Echelle spectra 
were reduced using standard procedures, giving complete wavelength coverage
of the entire visual spectral region at high S/N ($>$\,100) and high
resolution ($R$\,$\simeq$\,48\,000). 
\vskip3mm
\parbox{5.5cm}{
\tabcolsep=2.5pt
\begin{tabular}{ll}
\multicolumn{2}{c}{\parbox{5.4cm}{
~~~~{\bf Table 1.}{\ Non-LTE model atoms}}}\\
\tablerule
Ion & Source\\
\tablerule
H               & Przybilla \& Butler~(2004)\hhuad\\[-1pt]
He\,{\sc i/ii}  & Przybilla~(2005)\hhuad\\[-1pt]
C\,{\sc ii/iii} & Nieva \& Przybilla (in prep.)\hhuad\\[-1pt]
N\,{\sc ii/iii} & Przybilla \& Butler~(2001),\hhuad\\[-1pt]
                & with extensions\hhuad\\[-1pt]
O\,{\sc ii}     & Becker \& Butler~(1988)\hhuad\\[-1pt]
Mg\,{\sc ii}    & Przybilla et al.~(2001)\hhuad\\[-1pt]
S\,{\sc ii/iii} & Vrancken et al.~(1996),\hhuad\\[-1pt]
                & with updated atomic data\hhuad\\[-1pt]
\tablerule
\end{tabular}
}
\hfill
\parbox[b]{5.5cm}{
\tabcolsep=1.5pt
\begin{tabular}{lrr}
\multicolumn{3}{c}{\parbox{5.4cm}{
~~~~{\bf Table 2.}{\ Stellar Parameters}}}\\
\tablerule
 & HD\,205805 & Feige\,49\hhuad\\[-1pt]
\tablerule
$T_{\rm eff}$\,(K)         & 25\,000 & 35\,000\hhuad\\[-1pt]
$\log g$\,(cgs)            &    5.00 &    5.25\hhuad\\[-1pt]
$\xi$\,(km/s)              &       0 &       2\hhuad\\[-1pt]
$y$                        &    0.01 &    0.15\hhuad\\[-1pt]
$v \sin i$\,(km/s)         &       0 &       0\hhuad\\[-1pt]
\tablerule
\rule{0cm}{1.47cm}
\end{tabular}
}
\vskip2mm

\sectionb{3}{STELLAR PARAMETERS}

Standard methods are used for the stellar parameter determination, by
simultaneous fitting the hydrogen Balmer and helium line profiles, and when possible
utilising~the He\,{\sc i/ii} ionization balance. We deviate from 
the usual approach based on $\chi^2$-fitting on grids of synthetic spectra, because of
the largely increased compu-

\vbox{
\centerline{\psfig{figure=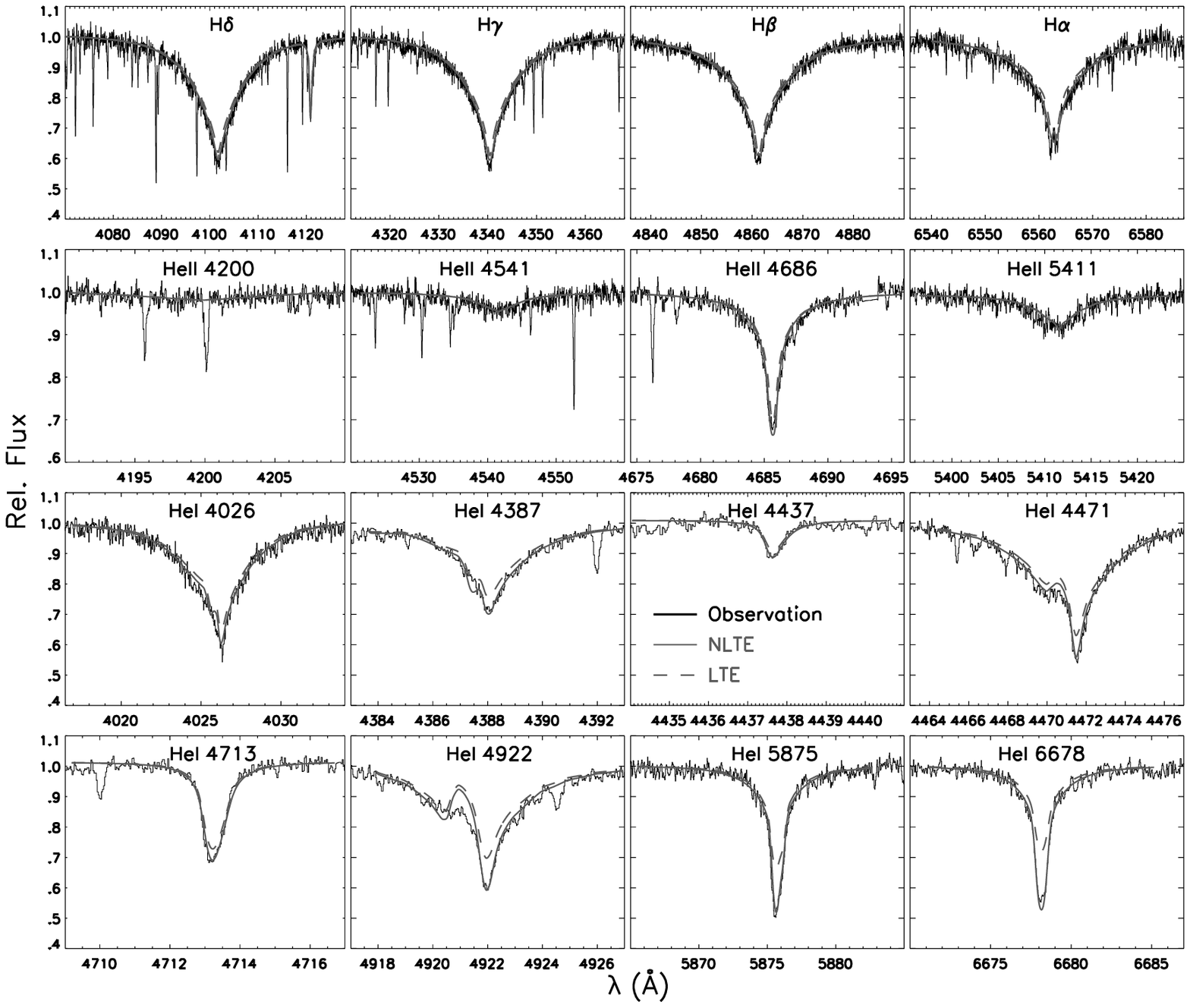,width=120truemm,clip=}}
\vspace{-5mm}
\captionc{1}{Fits to H and He lines in the helium-rich sdB Feige\,49}
}
\vskip5mm

\noindent tational expenses for OS model atmospheres and
non-LTE line formation. For the moment, the fits are done by eye. The
resulting stellar parameters, effective temperature $T_{\rm eff}$,
surface gravity $\log g$, microturbulence $\xi$, helium
abundance $y$ (by number) and projected rotational velocity $v\sin i$, 
are summarised in Table~2. 

Examples for the quality of the modelling achievements are displayed in
Figures~1 and 2, where
our spectrum synthesis for several diagnostic features (grey lines) is compared 
to observation (histogram) for the two sample stars.
Obviously, non-LTE line formation (full lines) improves enormously on the LTE
spectrum synthesis for identical stellar parameters (dashed lines), 
resulting in a practically perfect match of theory and observation (except
for the forbidden component of He\,{\sc i}\,$\lambda$4922). 

The largest discrepancies between non-LTE and LTE
results occur for some He\,{\sc i} lines. Non-LTE equivalent widths are
larger by up to $\sim$35\% in particular for the strong features in the red.
Both, line wings and cores can be affected. This is because the lower levels of the
transitions (for principal quantum number $n$\,$=$\,2) are overpopulated
relative to states at higher excitation energy that couple closely to the
He\,{\sc ii} ground state, which is in detailed balance. Among the 
He\,{\sc ii} lines only He\,{\sc ii} $\lambda$4686 gets slightly strengthened by
non-LTE effects. A similar situation as with the He\,{\sc i} lines occurs
with the hydrogen Balmer lines because of an analogous non-LTE overpopulation of the 
$n$\,$=$\,2 level, which also can affect the line wings (most notable for
H$\alpha$ in Feige\,49).
The non-LTE effects get reduced with decreasing temperature. In HD\,205805
only the line cores are affected.

\vbox{
\centerline{\psfig{figure=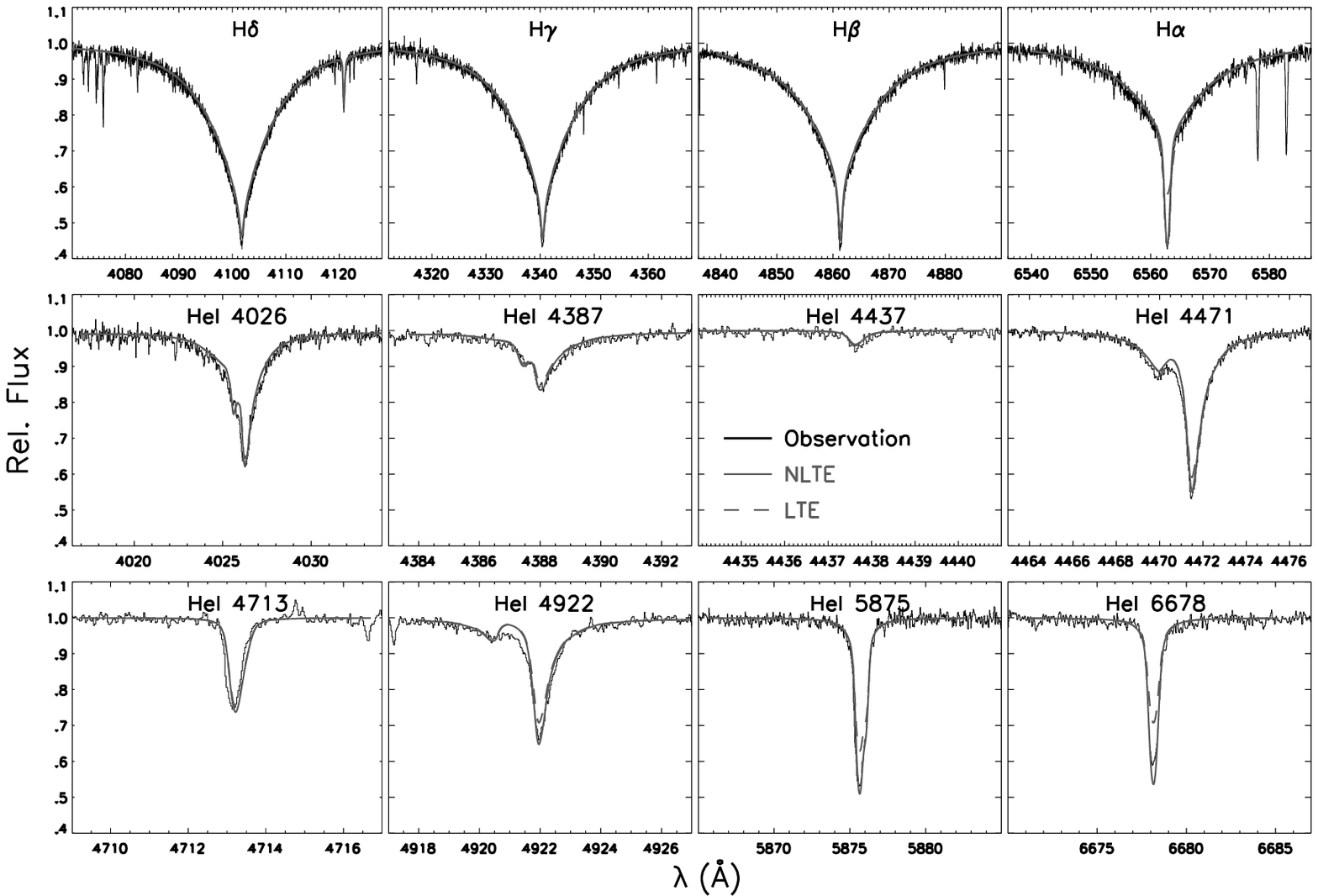,width=120truemm,clip=}}
\vspace{-2mm}
\captionc{2}{Fits to H and He lines in the helium-poor sdB HD\,205805}
}
\vskip5mm

It is obvious that the overall differences in the fitting process
will impact the stellar parameter determination. In the present case this
transfers into a $\Delta T_{\rm eff}$ of $+$1\,000 and $+$1\,200\,K and a
$\Delta \log g$ of $+$0.10 and $+$0.12\,dex for Feige\,49 and HD\,205805
relative to the LTE results of Edelmann et al. (these proceedings). These 
{\it systematic} shifts are slightly larger than the typically attributed uncertainties of
sdB analyses. Further refinements can be expected from the use of (multiple) 
non-LTE ionization equilibria of metals which are even more sensitive to
stellar parameter changes than the hydrogen and helium lines. So far, only the 
S\,{\sc ii/iii} ionization equilibrium has been used to verify the parameter
determination for HD\,205805 from H and He line profile fits (see below).

\sectionb{4}{ELEMENTAL ABUNDANCES}

Besides unbiased stellar parameters, which are a prerequisite for meaningful
comparisons with stellar evolution computations, also the surface abundances
of the heavier elements bear important information. They allow to put
observational constraints on formation/evolution scenarios of the stars and 
on transport processes (in particular diffusion for sdBs) in stellar atmospheres.  

For the time being, non-LTE abundances are determined only for H, He, C, N, O and the
$\alpha$-elements Mg and S because we lack realistic non-LTE model atoms 
for the other chemical species. The lighter elements are of interest because
of their involvement in fusion reactions, either as catalysts or as burning
products, thus giving clues on the nature of the sdB progenitors. The
$\alpha$-process elements on the other hand can be used as tracers for the 
stellar metallicity. All the present-day abundances may be of course subject
to diffusion. We intend to extend the study to Al, Si and Fe
in the near future. Work on this is in progress.

\vbox{
\centerline{\psfig{figure=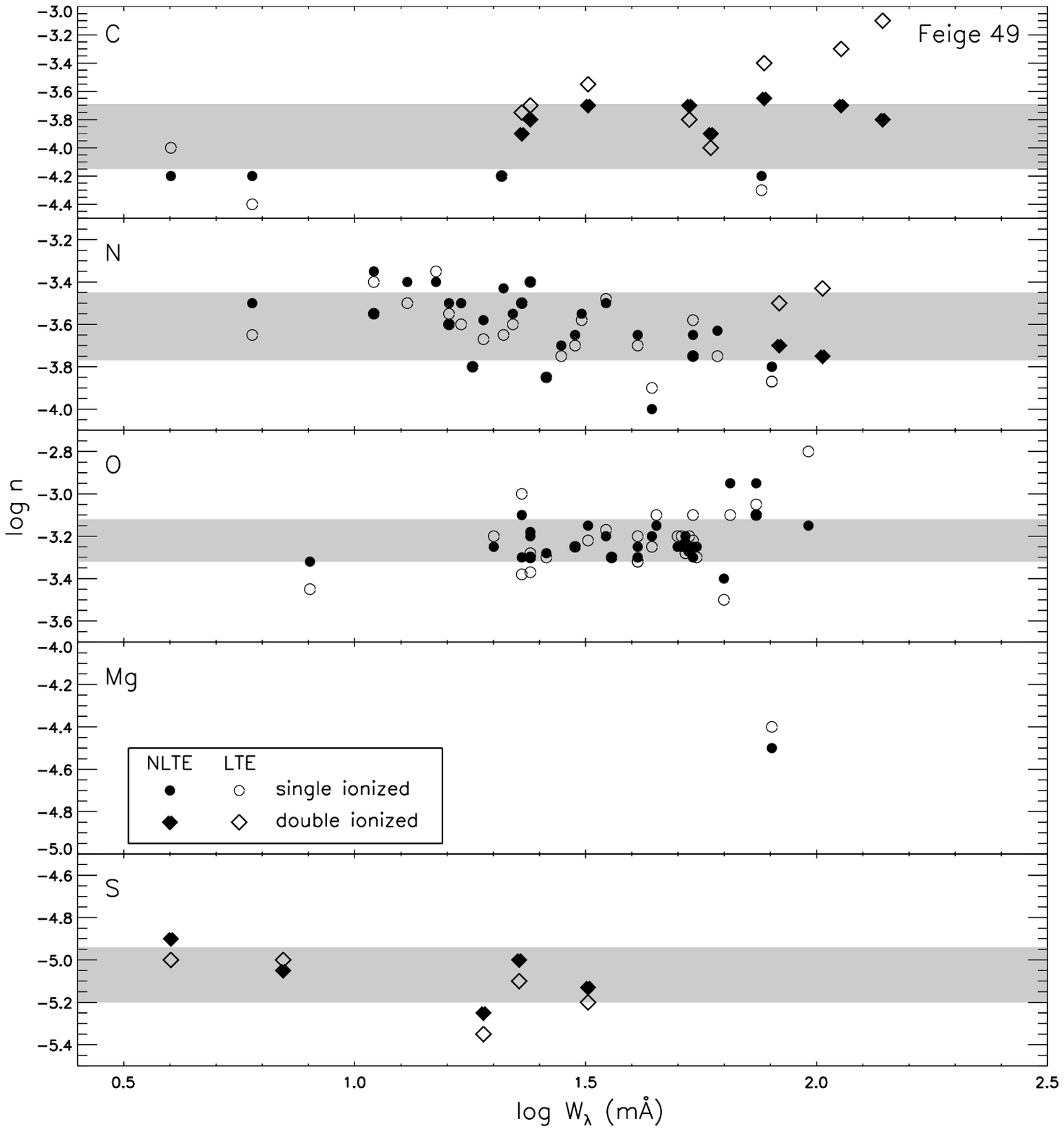,width=120truemm,clip=}}
\vspace{-3mm}
\captionc{3}{Elemental abundances in Feige\,49 from individual spectral lines}
}
\vskip3mm

Elemental abundances are derived from line-profile fits, 
using a $\chi^2$-minimisation technique
based on small grids of synthetic spectra with varying abundances for given
stellar parameters.
This puts tighter constraints than the standard
equivalent-width analysis. Abundances (by number) from individual spectral lines as a
function of equivalent width are displayed in Figures~3 and 4. Non-LTE
abundances are denoted by full and LTE results by open symbols;
circles mark single-ionized and diamonds double-ionized species. The grey
bands indicate the 1$\sigma$-uncertainty range of the resulting abundances 
for the chemical species. 

The results have to be viewed as preliminary, as abundances from the different 
ionization stages of the elements -- in particular C\,{\sc ii/iii} and
N\,{\sc ii/iii} -- indicate a need to improve on the stellar
parameters. Metal ionization equilibria react much more sensitively to changes 
than the hydrogen and helium lines, such that the necessary fine-adjustments 
will barely impact the fit quality of the latter. However, already now the
advantages of non-LTE computations become apparent: a tendency towards a
reduced statistical scatter relative to LTE and a reduction~of

\vbox{
\centerline{\psfig{figure=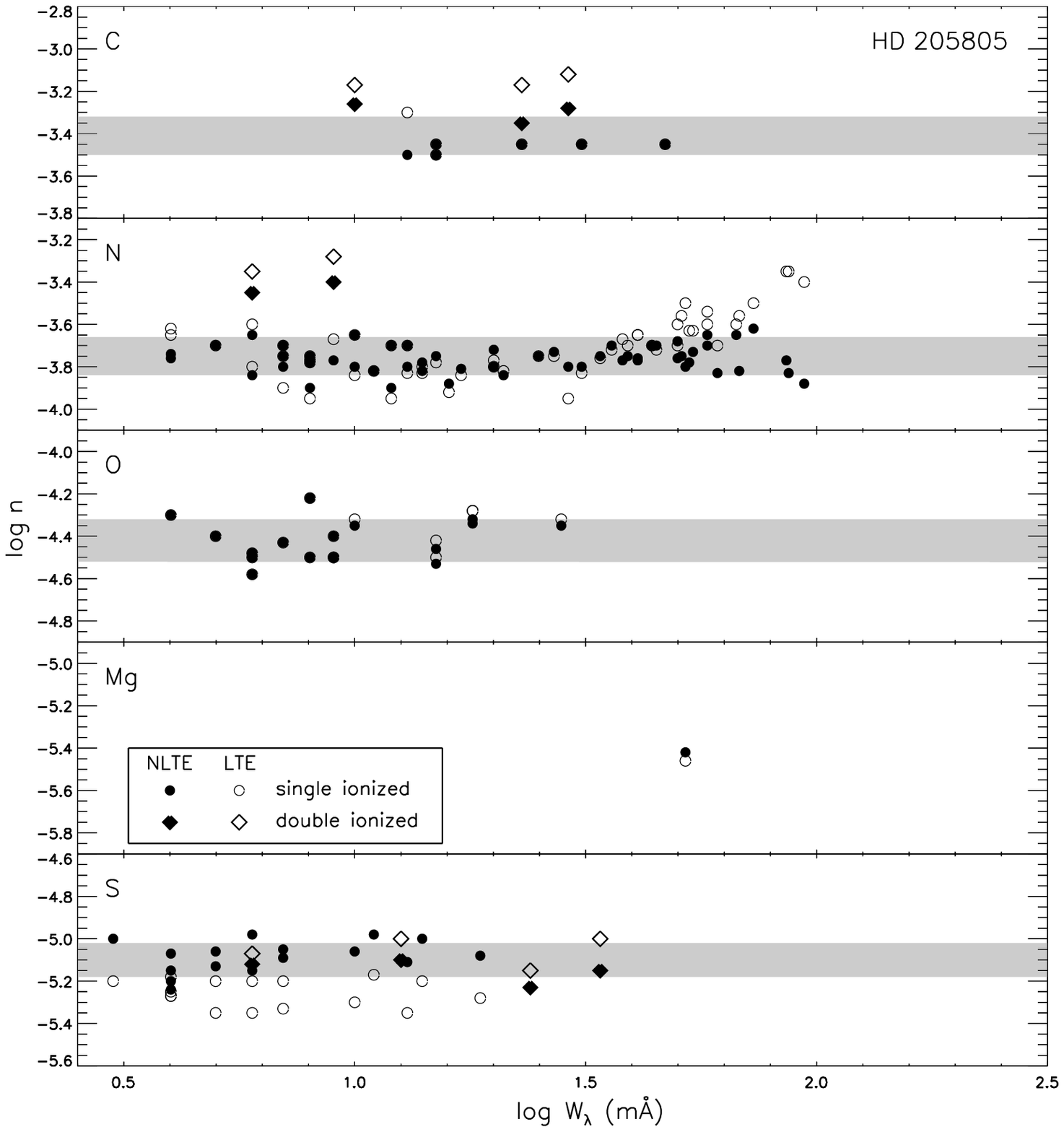,width=120truemm,clip=}}
\vspace{-3mm}
\captionc{4}{Elemental abundances in HD\,205805 from individual spectral lines}
}
\vskip8mm

\begin{center}
\vbox{\norm\small
\tabcolsep=3pt
\begin{tabular}{lrrrrrrrrr}
\multicolumn{10}{c}{\parbox{101mm}{
{\bf \ \ Table 3.}{\ Metal abundances in the sample stars, relative to the
solar standard (Grevesse \& Sauval~1998): 
$[\log\,n]\,=\,\log\,n_{\star}\,-\,\log\,n_{\odot}$}}}\\
 \tablerule
           &      &   C\,{\sc ii} & C\,{\sc iii} & N\,{\sc ii} & N\,{\sc iii} & O\,{\sc ii} & Mg\,{\sc ii} & S\,{\sc ii} & S\,{\sc iii}\\
\tablerule   
Feige\,49  & NLTE & $-$0.72 & $-$0.29 & $+$0.48 & $+$0.35 & $-$0.05 & $-$0.08  & {\ldots} & $-$0.27\\
           &      & {\ldots}&    0.10 &    0.16 &    0.04 &    0.10 & {\ldots} & {\ldots} &    0.13\\
           & LTE  & $-$0.75 & $-$0.10 & $+$0.44 & $+$0.61 & $-$0.06 & $+$0.02  & {\ldots} & $-$0.33\\
           &      &    0.17 &    0.30 &    0.15 &    0.05 &    0.14 & {\ldots} & {\ldots} &    0.15\\[2mm]
HD\,205805 & NLTE & $+$0.01 & $+$0.18 & $+$0.32 & $+$0.65 & $-$1.25 & $-$1.00  & $-$0.28 & $-$0.35\\
           &      &    0.03 &    0.05 &    0.06 &    0.04 &    0.10 & {\ldots} &    0.08 &    0.06\\
           & LTE  & $+$0.05 & $+$0.33 & $+$0.38 & $+$0.77 & $-$1.24 & $-$1.04  & $-$0.46 & $-$0.25\\
           &      &    0.07 &    0.03 &    0.14 &    0.05 &    0.10 & {\ldots} &    0.07 &    0.07\\
\tablerule
\end{tabular}
}
\end{center}

\vbox{
\centerline{\psfig{figure=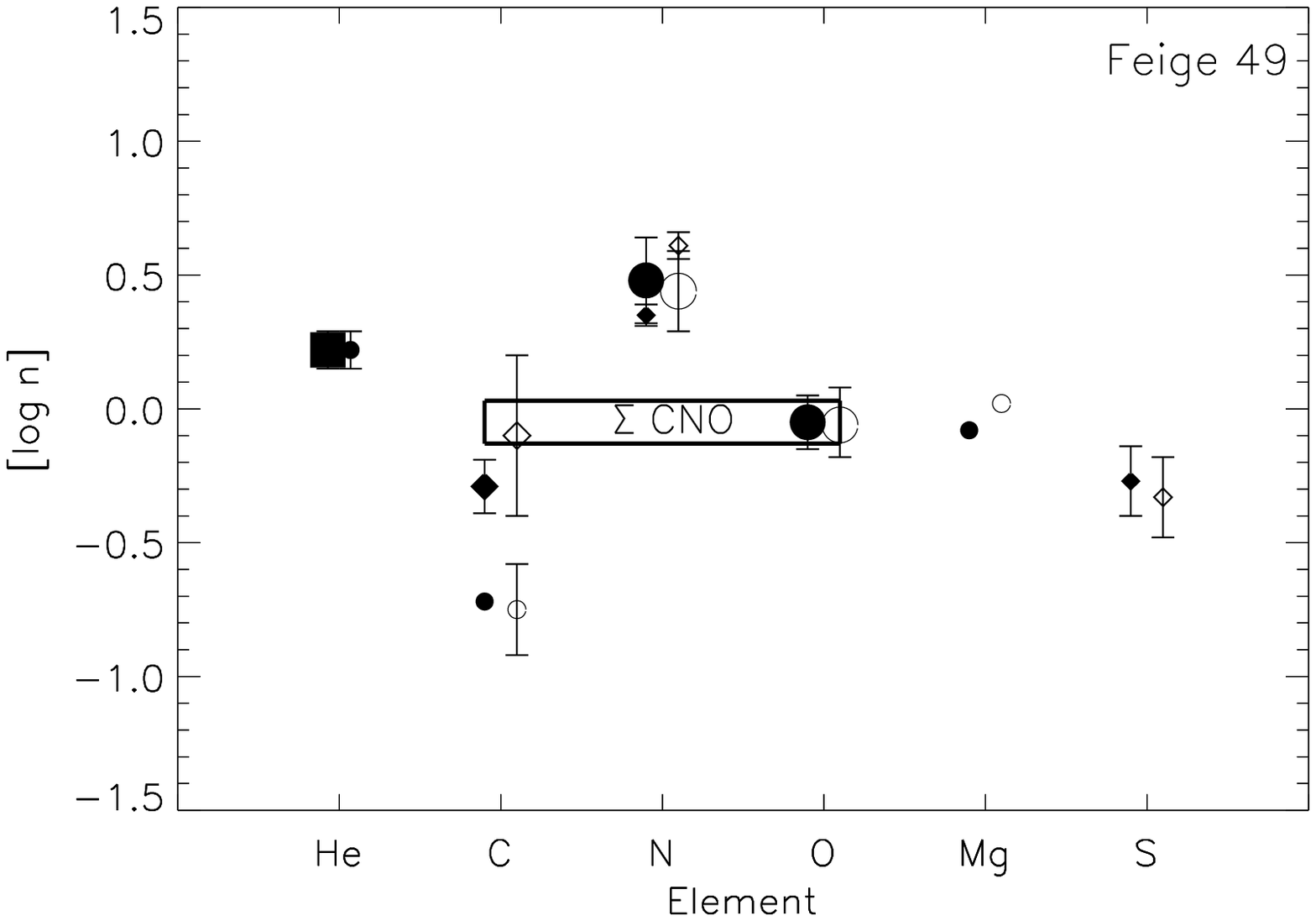,width=81mm,clip=}}
\vspace{-4mm}
\centerline{\psfig{figure=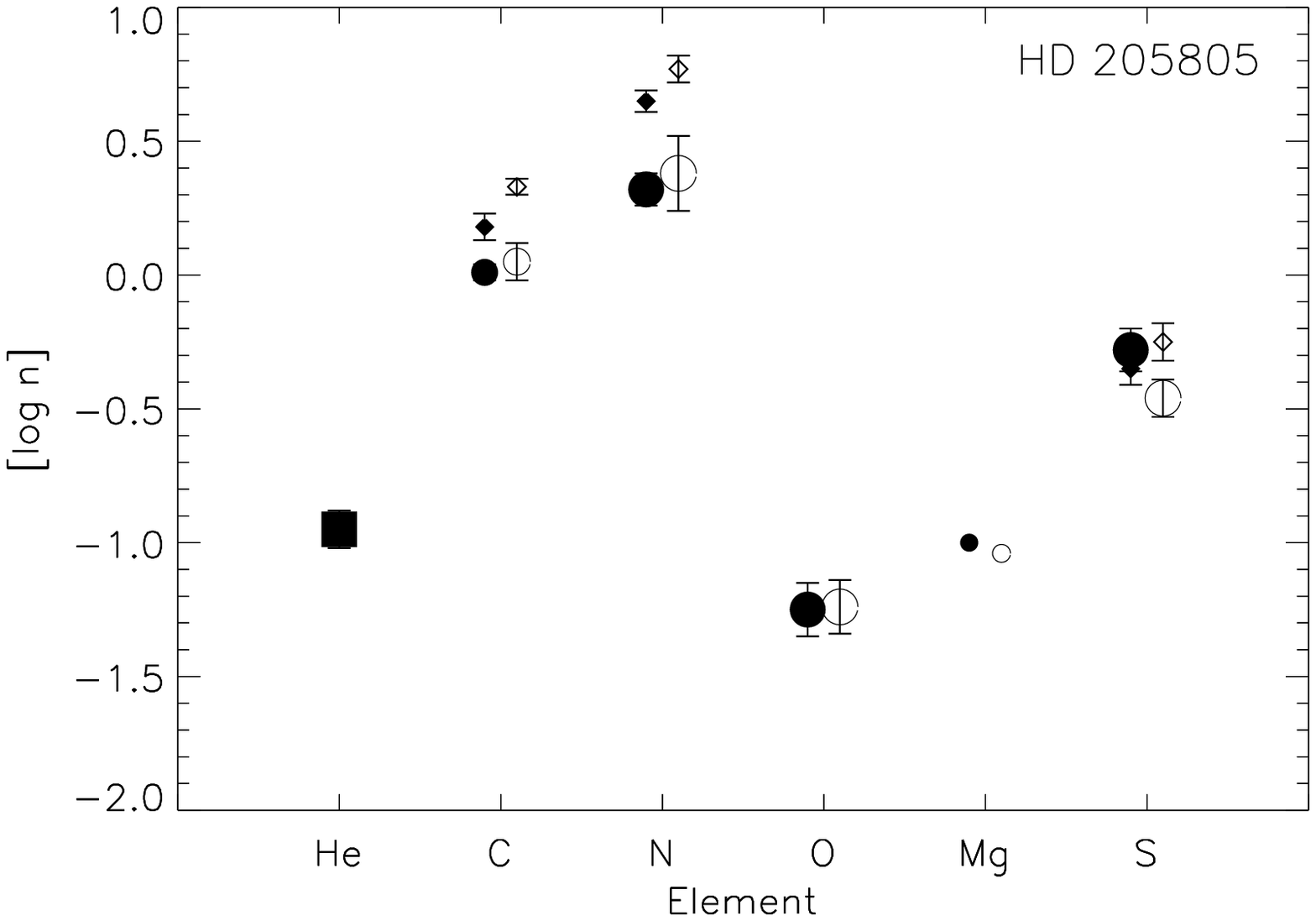,width=81mm,clip=}}
\vspace{-3mm}
\captionc{5}{Elemental abundances in the two sample sdBs (symbols as in Fig.~4)}
}
\vskip5mm

\noindent systematic trends, like for N\,{\sc ii} in HD\,205805.~The~aim is 
to push the statisti\-cal 1$\sigma$-uncertainties below 0.1\,dex, like in
similar computations for BA-type supergiants (Przybilla~2002) and B-type main 
sequence stars 
(Nieva \& Przy\-billa,~in~prep.). The systematic uncertainties need to be
determined, but they can be expected to be of the order $\sim$0.1\,dex on
the mean, as in the other cases.

The metal abundances of the sample stars relative to the solar standard are
summarised in Table~3. For each ionic species non-LTE and LTE results with
uncertainties derived from the line-to-line scatter are given. The data is
also displayed in Figure~5, using the same symbols as in the figures before,
with the symbol size encoding the number of lines used for the abundance determination.

The abundance pattern for Feige\,49 indicates mixing of the surface layers
with CNO-processed material: He and N are enriched and C is depleted. 
Oxygen appears to be unchanged relative to the other $\alpha$-elements, which
indicate approximately solar metallicity for this star. Note that the sum of
the CNO abundances also gives a value close to solar, emphasising the
catalyst r\^ole of these species. 
The pattern is unusual for a sdB star.
On the other hand, HD\,205805 shows an abundance pattern more typical for
sdB stars, indicating that diffusion processes have been active. 

\sectionb{4}{CONCLUSIONS \& PROSPECTS}

Hybrid non-LTE analyses of two sdB stars demonstrate that non-LTE effects
have to be accounted for in the stellar parameter and abundance determination,
resulting in excellent fits of {\it all} the diagnostic spectral features.
Systematic shifts in the basic parameters $T_{\rm eff}$ and $\log g$ are
implied and non-LTE abundance corrections on the order of
$\sim$0.05--0.25\,dex on the mean. The latter are similar to corrections of
metal abundances in (less-luminous) BA-type {\it supergiants} (Przybilla~2002).

The next steps in the refinements of the method will include the necessary
fine-adjustments in the stellar parameter determination in order to meet the
constraints imposed by various metal ionization equilibria. The
uncertainties in the metal abundances can then be expected to drop below
0.1\,dex, as experience tells us from analyses of objects in other parts of
the Hertzsprung-Russell diagram. Additional non-LTE model atoms are required for a few other
important diagnostic chemical species, like aluminium (an element with odd
neutron number), silicon (another $\alpha$-element, highly valuable for
stellar parameter estimations) and iron (to cover iron group abundances). 

Similar analyses of a larger sample of sdB stars, and an extension towards
the sdO regime, will allow for unbiased positioning in the 
$T_{\rm eff}$--$\log g$ plane. This will help to delineate the different
populations of subluminous stars with unprecedented accuracy and may provide
the crucial clues for uncovering their formation history 
(single vs. binary star evolution channels). A stellar sample with highly
accurate parameter and abundance determinations will hereby also put tight observational
constraints on the stellar evolution computations. These can be used for an
empirical calibration of the parameterisation of complex (hydrodynamical)
phenomena involved in the stellar evolution calculations, thus leading to an
improved modelling.

\References

\refb
Becker~S.~R., Butler~K. 1988, A\&A, 201, 232

\refb
Butler~K., Giddings~ J.~R.~1985, in Newsletter on Analysis of
Astronomical Spectra, No.\,9, Univ. London

\refb
Giddings~J.~R. 1981, Ph.\,D. thesis, Univ. London

\refb
Grevesse~N., Sauval~A.~J. 1998, Space Sci. Rev., 85, 161

\refb
Heber U., Edelmann H. 2004, Ap\&SS, 291, 341

\refb
Kaufer A., Stahl O., Tubbesing S., et al.~1999, ESO Messenger, 95, 8

\refb
Kurucz~R.~L. 1996, in {\it Model Atmospheres and Spectrum Synthesis}, eds.
S.~J.~Adelman, F.~Kupka, \& W.~W.~Weiss, ASP Conf. Ser., 108, 160

\refb
Przybilla~N. 2002, Ph.\,D. thesis, Univ. Munich

\refb
Przybilla~N. 2005, A\&A, 443, 293

\refb
Przybilla~N., Butler~K. 2001, A\&A, 379, 955

\refb
Przybilla~N., Butler~K. 2004, ApJ, 609, 1181

\refb
Przybilla~N., Butler~K., Becker~S.~R., Kudritzki~R.~P. 2001, A\&A, 369, 1009

\refb
Stehl\'e~C., Hutcheon~R. 1999, A\&AS, 140, 93

\refb
Vrancken~M., Butler~K., Becker~S.~R. 1996, A\&A, 311, 661

\end{document}